\documentclass[dvips,twocolumn,aps,prl,nofootinbib]{revtex4}
\usepackage{graphicx}
\usepackage{todonotes}

\newcommand{\lapl}[1]{\ensuremath{\big\langle #1 \big\rangle_\omega}}
\newcommand{\avg}[1]{\ensuremath{\big\langle #1 \big\rangle}}

\renewcommand{\vec}[1]{\ensuremath{\mathbf{#1}}}
\newcommand{\bsym}[1]{\boldsymbol{#1}}
\usepackage{amsmath,amssymb,amsfonts,mathrsfs}
\begin{document}
\frenchspacing
\newcommand{\preprintclearpage}{\clearpage}

\author{M.~Sega}
\email{marcello.sega@univie.ac.at}
\affiliation{Institut f\"ur Computergest\"utzte Biologische Chemie, University of Vienna, W\"ahringer Strasse 17, 1090 Vienna, Austria}
\author{S.S.~Kantorovich}
\affiliation{Faculty of Physics, University of Vienna, Boltzmanngasse 5, 1090 Vienna, Austria}
\affiliation{Ural Federal University, Lenin av. 51, 620083 Ekaterinburg, Russia}
\author{C.~Holm}
\author{A.~Arnold}
\affiliation{Institut f\"ur Computerphysik, Universit\"at Stuttgart, Allmandring 3, 70569 Stuttgart, Germany}

\title{Kinetic and ion pairing contributions in the dielectric spectra of electrolyte aqueous solutions}

\begin{abstract}
Understanding dielectric spectra can reveal important information
about the dynamics of solvents and solutes from the dipolar relaxation
times down to electronic ones.  In the late 1970s, Hubbard and
Onsager predicted that adding salt ions to a polar solution would
result in a reduced dielectric permittivity that arises from the
unexpected tendency of solvent dipoles to align opposite to the
applied field.  So far, this effect has escaped an experimental
verification, mainly because of the concomitant appearance of
dielectric saturation from which the Hubbard-Onsager decrement
cannot be easily separated.  Here we develop a novel non-equilibrium
molecular dynamics simulation approach to determine this decrement
accurately for the first time. Using a thermodynamic consistent
all-atom force field we show that for an aqueous solution containing
sodium chloride around 4.8 Mol/l, this effect accounts for 12\% of
the total dielectric permittivity. The dielectric decrement can be
strikingly different if a less accurate force field for the ions
is used. Using the widespread GROMOS parameters, we observe in fact
an {\it increment} of the dielectric permittivity rather than a
decrement.  We can show that this increment is caused by ion pairing,
introduced by a too low dispersion force, and clarify the microscopic
connection between long-living ion pairs and the appearance of
specific features in the dielectric spectrum of the solution.
\end{abstract}

\maketitle

%%%%%%%%%%%%%%%%%%%%%%%%%%%%%%%%%%%%%%%%%
%                <Volume>  
%  KBFF           54.06
%  GROMOS/SPC     56.60
%  KBFF2          53.92
%  GROMOS/SPCE    55.67
%%%%%%%%%%%%%%%%%%%%%%%%%%%%%%%%%%%%%%%%%
It is well known that addition of salt to a polar liquid changes
its dielectric properties, in particular by lowering its static
dielectric permittivity~\cite{kremer03a}. Much less is known about
the actual physical mechanisms which contribute to the decrement.
Hubbard and Onsager~\cite{hubbard77a} first proposed that the motion
of ions, dragged by an external electric field, should induce a
depolarization of the solvent molecules, by orienting their dipole
moments opposite to the electric field.  This effect would be
genuinely kinetic in nature, not having any static
equivalent\cite{wolynes80a}. However, the existence of this effect
has never been proven experimentally, because it is masked by the
decrement caused by dielectric saturation. In other words, solvent
molecules in the vicinity of an ion are strongly polarized by its
local field, and do not take part in the usual dielectric relaxation
process, therefore contributing to the reduction of the dielectric
permittivity.  Molecular dynamics simulations represent an ideal
tool to investigate this effect, but so far a coherent framework
for the calculation of the Hubbard-Onsager decrement has not been
proposed.  Here, we derive expressions for the dielectric decrement
suited for equilibrium and out-of-equilibrium simulations.  We use
the out-of-equilibrium approach to estimate with high accuracy the
static limit of the kinetic decrement, showing that there is indeed
a sizable kinetic contribution, even though the Hubbard-Onsager
continuum theory overestimates it.  In addition, we scrutinize
directly the impact of long-living ion pairs  on the dielectric
spectrum of the solution.  Finally, we show that the ionic current
correlation results in a dielectric increment.  This kinetic
contribution is often disregarded, but can counterbalance the
Hubbard-Onsager kinetic decrement, even in the absence of ion
pairing.

Several approaches have been proposed to date to compute the
dielectric spectrum $\epsilon(\omega)$ in presence of free
ions~(e.g.\,\cite{caillol86a,schroeder06c}). The common denominator
of their theoretical background is the separate coupling of the
dipole vector of neutral moieties and of the current of charged
ones to the electrostatic scalar and vector potentials, respectively.
As only the curl-free part of the electric field survives in classical
simulations, all these approaches are equivalent~\cite{sega13a} to
calculating the dielectric spectrum from that of the conductivity
$\sigma(\omega)$ as $ \epsilon(\omega) = 1  + 4\pi i
\sigma(\omega)/\omega,$ where the quantity $4 \pi i \sigma(\omega)/\omega
$ is sometimes called the generalized dielectric function and denoted
by $\Sigma(\omega)$.  The conductivity spectrum, in turn, can be
computed in the linear response regime from the time correlation
current\cite{kubo57a} $
\sigma(\omega)=\beta\lapl{\vec{j}_T(t)\vec{j}_T(0)}/3V $ of the
total electric current $\vec{j}_T$, generated by both partial charges
on solvent molecules and by ionic charges.  The symbols
$\left\langle\ldots\right\rangle$ and
$\left\langle\ldots\right\rangle_\omega$ denote the canonical average
and its Fourier-Laplace transform, $\langle f \rangle_\omega =
\int_0^\infty \langle f \rangle\exp(i \omega t)  \mathrm{d}t$,
respectively.

The total current  can be split arbitrarily into different
contributions. For aqueous solutions of atomic ions, however, 
it is natural to decompose it into solvent contributions, $\vec{j}_w$, and
ionic contributions, $\vec{j}_i$. The two currents stem from motion of the partial charges
of water molecules and from the ionic charges, respectively. This
splitting has a particularly important meaning in the context of
dynamical contributions. In the linear response relation
$\Delta\vec{j}_a(\omega)\propto\lapl{\vec{j}_a(t)\vec{j}_b(0)}=
C_{ab}(\omega)$  (where subscript $a$ and $b$ are completely general),
the current $\vec{j}_b$ is the derivative of the dipole moment that
couples to the perturbing field in the Hamiltonian. Hence, for
example, in an aqueous solution of ions, the correlation $C_{ww}$
represents the response of water in a fictitious case where only
water molecules - and not the ions - are coupled to the electric
field. Similar considerations apply to the other possible permutations
$C_{wi}$, $C_{iw}$, and $C_{ii}$.  Evidently the zero-frequency
limit of $\Delta \epsilon_{iw}(\omega) = 4\pi\beta i C_{wi}(\omega) /(
3\omega V )$ is precisely the first dynamic mechanism identified by Hubbard and Onsager, namely, the depolarization of water molecules due to the motion of ions. The symmetry between $C_{wi}$ and $C_{iw}$ is nothing
else but an instance of Onsager reciprocal relations, and shows
directly that the second dynamical mechanism identified by Hubbard and 
Onsager (namely, a reduction in ion conductivity due to the
rotation of water molecules) leads to the same contribution to
the static permittivity. The total kinetic decrement is therefore
$\Delta\epsilon_{HO}=2\Delta\epsilon_{wi}(0)$. The static dielectric permittivity can be written as $\epsilon=1+\Delta\epsilon_{wT}+\Delta\epsilon_{iT}=1+\Delta\epsilon_{ww}+\Delta\epsilon_{ii}+2\Delta\epsilon_{wi}$, where here and in the following the absence of a frequency dependency implies that the limit to zero frequency has to be taken.

\begin{table}
\begin{tabular}{lllll}
Model& atom & $\epsilon$ (kJ/mol) & $\sigma_{LJ}$ (nm) & charge ($e$)\\
\hline\hline
KBFF  & Na & 0.320$^1$ & 0.245 & { }1\\
      & Cl & 0.470     & 0.440 &   -1\\ 
GROMOS& Na & 0.062     & 0.258 & { }1\\
      & Cl & 0.446     & 0.445 &   -1\\ 
SPC/E & O  & 0.650     & 0.316 &   -0.8476\\
      & H  & 0.0       & 0.0   & { }0.4238\\
\end{tabular}
\begin{flushleft} {\footnotesize }
Geometric mixing rules for both $\sigma_{LJ}$ and $\epsilon$ are employed.\\
$^1$ The interaction energy between oxygen and sodium atoms, has to be scaled by a factor 0.75 in KBFF.\end{flushleft}
\caption{Lennard Jones interaction parameters}
\label{table1}
\end{table}

We have investigated the dielectric spectrum of aqueous solutions of
sodium chloride, performing extensive MD simulations.  The systems
were composed of 1621 water molecules (SPC/E potential\cite{berendsen87a})
and 160 ion pairs, corresponding to a salt concentration of about
4.7 -- 4.9 mol/l, depending on the average volume yielded by the
use of different potentials for the ions. Distinct simulations were
performed using the GROMOS\cite{scott99a} and the
KBFF\cite{weerasinghe03a} force fields (see Table~\ref{table1} for
the parameters).  The ensemble of choice is the isothermal/isobaric
one at 298 K and 1 atm, realized using the
Nos\`e--Hoover\cite{nose84a,hoover85a} and Parrinello--Rahman
coupling schemes\cite{parrinello81a}.  The electrostatic interaction
has been computed using the Particle Mesh Ewald method in its smooth
variant\cite{essmann95a} with tin-foil boundary conditions, 4th
order spline interpolation on a grid with 0.12 nm spacing and a
relative strength of the interaction at 0.9 nm of $10^{-5}$.  Both
the Lennard-Jones and short-range part of the electrostatic potentials
were switched smoothly to zero using a fourth-degree polynomial in
the range from 0.9 to 1.2 nm.

Several 8 ns long runs with different initial conditions were performed,
up to a cumulative time of about 4 $\mu$s for each system. Water
and ionic currents were saved to disk every time step (1 fs) for
off-line analysis.  Current correlations were calculated on the
complete 8 ns datasets using a fast Fourier transform based
approach~\cite{futurelle71a,allen87a,press92a}. The correlations
computed on each dataset were eventually averaged, and spectra were
calculated using delays of up to 2 ns of the averaged correlations, as
at longer delay times the noise in the correlation degraded the quality
of the spectrum noticeably.
No tapering or padding procedure has been used in order to avoid any bias in the spectral features, especially in those
at low frequencies.

\begin{figure}[t]
\begin{center}
\includegraphics[bb=55 25 430 322,clip,width=\columnwidth]{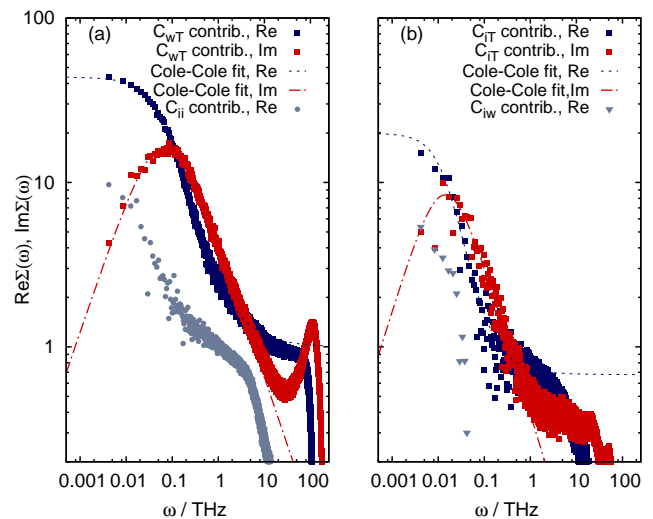}
\end{center}
\caption{Contributions to the dielectric spectrum in the GROMOS potential case. The water (left panel) an ions (right panel) contributions (squares), and Cole-Cole fit to their real (dashed line) and imaginary (dot-dashed) parts. In addition, the real part of the ion-ion self correlation is shown in the left panel (circles) and the water-ion cross correlation  is shown in the right panel (triangles).
\label{fig:gromos} }
\end{figure}

\begin{figure}[!t]
\begin{center}
\includegraphics[bb=55 25 430 322,clip,width=\columnwidth]{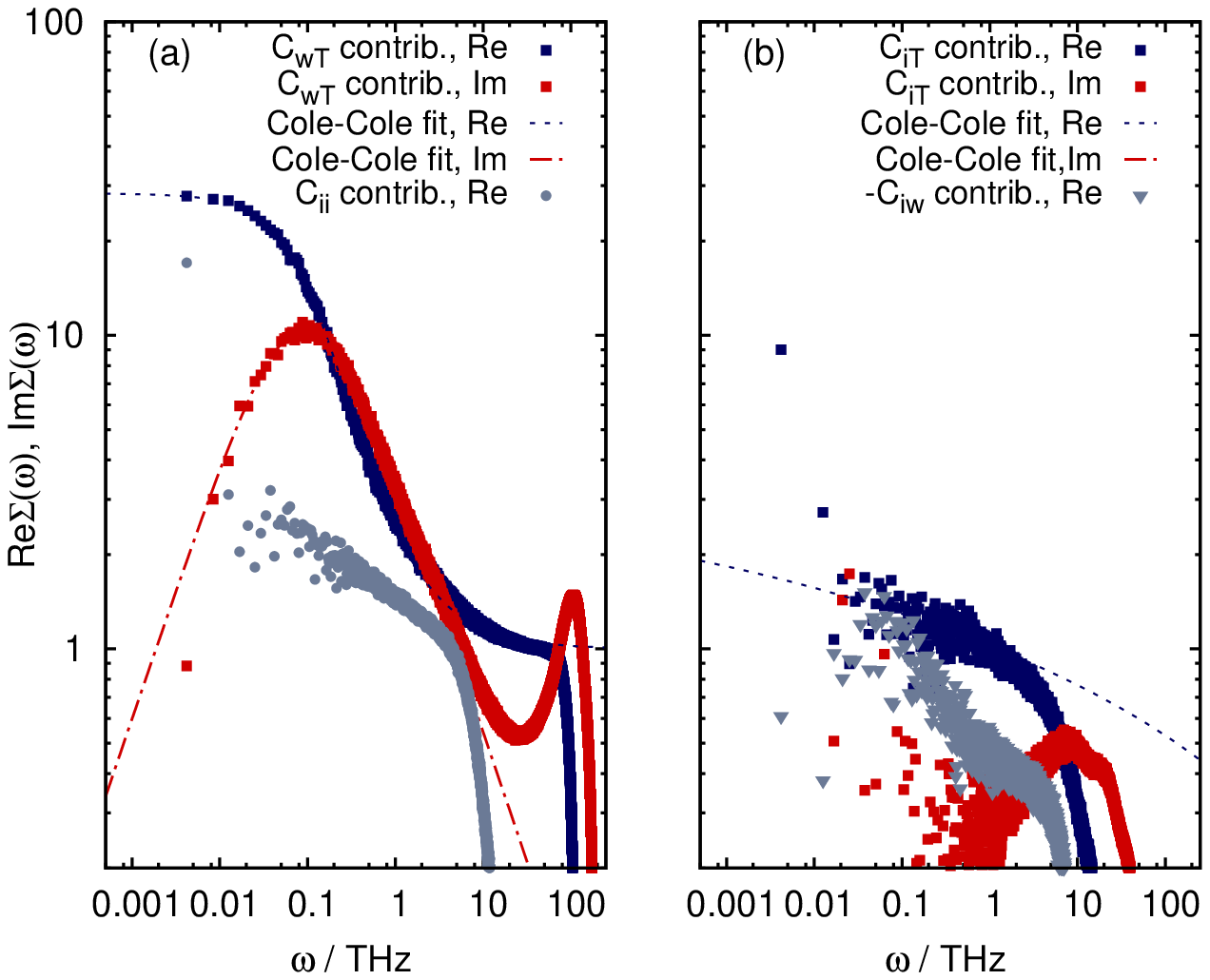}
\end{center}
\caption{Contributions to the dielectric spectrum in the KBFF potential case. The water (left panel) an ions (right panel) contributions (squares), and Cole-Cole fit to their real (dashed line) and imaginary (dot-dashed) parts. I addition, the real part of the ion-ion self correlation is shown in the left panel (circles) and the negative part of the water-ion cross correlation is shown in the right panel (triangles). 
\label{fig:kbff} }
\end{figure}

At first, we investigated the solution using
the GROMOS force field for Na and Cl ions~\cite{scott99a}, obtaining
the spectra reported in Fig.\ref{fig:gromos}. The spectrum has been
split into the contributions coming from water $C_{wT}$ and from
the ions $C_{iT}$, where the index $T$ again denotes the total current. In addition, the positive part of the cross correlations
$C_{wi}$ has also been reported.
A clear, main relaxation peak of water (left panel of Fig.\ref{fig:gromos})
appears at $\omega\simeq78$ GHz ($\tau=14.6$ ps), but the ionic contribution
shows the appearance of an important, unexpected relaxation process at $\omega\simeq14$
GHz ($\tau=70$ ps, d.c.~contribution subtracted).
A simultaneous fit of the real and imaginary part of the spectra
to the phenomenological Cole-Cole functional form augmented by a conductivity term,
\begin{equation}
\epsilon(\omega)=\epsilon_\infty+ \frac{\Delta\epsilon}{1+(i\omega\tau)^{\alpha}}+\frac{4\pi i}{\omega}\sigma,
\end{equation}
led to the best fit parameters (amplitude $\Delta\epsilon$, relaxation time
$\tau$ exponent $\alpha$ and conductivity $\sigma$) reported in Tab.\ref{table2}. 

\begin{table}[t]
\begin{tabular}{lccc}
                   & GROMOS & KBFF & KBFF+pairs \\
\hline
\hline
$\Delta\epsilon_{wT}$   &      $  42.1 \pm 0.3   $&     $ 27.4 \pm 0.1  $ &  $ 39.9\pm0.3   $  \\
$\Delta\epsilon_{iT}$   &      $  18.0 \pm 2.0   $&     $     -         $ &  $ 36.0\pm4.0   $  \\
$\Delta\epsilon_{wi}$             &      $  6.0^\dagger   $&     $ -1.7\pm 0.1^\ddagger $ &  $ 5.9^\dagger $  \\
%%%$\Delta\epsilon_{HO,mol}$         &      $                 $&     $ -0.14         $ &  $ -            $         \\
$\Delta\epsilon_{HO,cont}$        &      $  -6.2           $&     $ -8.0          $ &  $ -8.6         $         \\
$\tau_{wT}$ / ps                  &      $  14.6 \pm 0.2   $&     $ 10.6 \pm 0.1  $ &  $   17.1\pm0.3 $  \\
$\tau_{iT}$ / ps                  &      $   70.0 \pm 1     $&     $ -             $ &  $  110\pm10     $  \\
$\alpha_{wT}$                     &      $ 0.820 \pm 0.005 $&     $ 0.830\pm0.003 $ &  $0.800\pm0.005 $  \\
$\alpha_{iT}$                     &      $  0.91 \pm 0.01  $&     $ -             $ &  $0.94\pm0.01 $  \\
%%$\sigma $ / ps$^{-1}$          &      $ 0.049           $&     $ 0.090         $ &  $  0.058       $  \\
$\sigma$ / (S/m)                  &      $ 5.1\pm 0.1            $&     $ 10.2 \pm0.1   $ &  $  5.4\pm0.1       $  \\

\hline
\multicolumn{4}{l}{$^\dagger$ value of $\Sigma(\omega)$ at the smallest frequency;}\\
\multicolumn{4}{l}{$^\ddagger$ from NEMD simulation}
\end{tabular}
\caption{Several spectral properties.
\label{table2}
}
\end{table}

The contribution coming from the cross-correlation $C_{wi}$
is large and positive, in open contrast to the prediction of a
kinetic \emph{decrement} by the Hubbard--Onsager mechanism.
In the continuum approximation, i.\,e.\,, in the limit of infinite dilution and large ionic radius, this decrement can be estimated~\cite{hubbard79a} as
$\Delta\epsilon_{HO,cont} = -8 \pi \tau_{wT} \sigma (\epsilon_0
-\epsilon_\infty)/ 3 \epsilon_0 $, where $\sigma$,
$\epsilon_0$ and $\epsilon_\infty$ are the static conductivity,
static permittivity and infinite frequency dielectric permittivity
of the solution, and $\tau_{wT}$ is the (Debye-like) relaxation time
of the solvent.
In the ionic contribution to the spectrum (right panel of
Fig.~\ref{fig:gromos}), however, a feature which is compatible with
a Cole-Cole process with a characteristic time of about 70 ps is
present. This suggests that ion pairing is taking place on this
relatively large timescale  which is consisten with earlier reports
that the effective attraction between solvated Na and Cl ions in
the GROMOS force field is too strong~\cite{hess06b}.
Sodium chloride, instead, is a strong electrolyte which is generally
believed not to lead to ion pairing on such a timescale~\cite{buchner08a}.

To test our hypothesis, we simulated the NaCl solution using the
Kirkwood--Buff force field (KBFF)~\cite{weerasinghe03a}, which is
more accurate than the GROMOS one in reproducing structural and energetic properties
of solvated sodium chloride. The water contribution 
is qualitatively similar to the GROMOS case, but the ion contribution differs dramatically.
The prominent ion relaxation process at about 14 GHz (70 ps), which characterized
the GROMOS case, disappears completely, and the zero-frequency
contribution becomes so small that it can hardly be estimated.

\begin{figure}[!t]
\begin{center}
\includegraphics[bb=55 25 430 322,clip,width=\columnwidth]{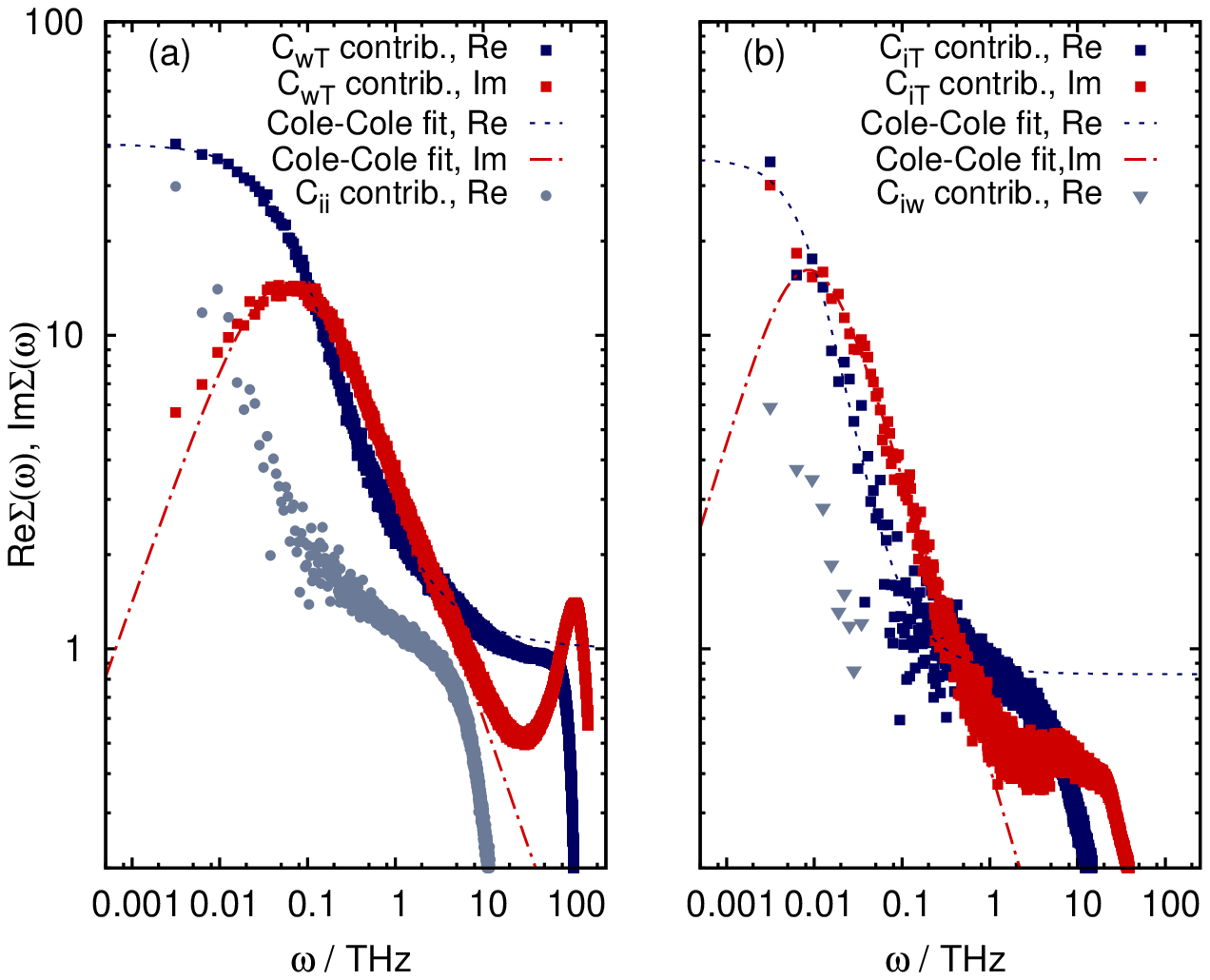}
\end{center}
\caption{Contributions to the dielectric spectrum in the KBFF potential case, when ion pairs are enforced. The water (left panel) an ions (right panel) contributions (squares), and Cole-Cole fit to their real (dashed line) and imaginary (dot-dashed) parts. I addition, the real part of the ion-ion self correlation is shown in the left panel (circles) and the water-ion cross correlation  is shown in the right panel (triangles).
\label{fig:kbff2} }
\end{figure}

To check explicitly that the phenomenon underlying the relaxation
observed in the GROMOS case is ion pairing, we performed another
series of simulations using the KBFF parameters. This time, however,
we created artificial ion pairs out of 50\% of the ions in solution
by introducing a stiff bond between the Na and Cl ions with a bond
length corresponding to the first peak in the radial distribution
function of the non-bonded case. The effect of persistent pairing
on the dielectric relaxation of ions modeled using KBFF can be
appreciated in Fig.\ref{fig:kbff2}.  A relaxation process appears
again at low frequencies ($\omega\simeq 9$ GHz, $\tau=110$ ps), qualitatively very similar to the one
observed using GROMOS force field, a clear indication that the
process has to be attributed to ion pairing.  This also shows, in
particular, that it is the contact ion pair mechanism which is
involved in the present case,  as the first peak of the radial
distribution function is located slightly closer than 0.3 nm, and
therefore a water molecule can not be accommodated in between the
two ions.  In this respect, KBFF proves to be superior to GROMOS
not only in describing the structural properties, but also the
dynamical ones, as it correctly reproduces the absence of ion pairs
that are stable on time scales longer than 10-100 ps.

Nevertheless, the failure of the GROMOS force field in reproducing
the spectrum of aqueous NaCl solutions has provided new important
knowledge.  On the one hand, this is the first direct observation
of the relaxation process associated with ion pairing, with a
characteristic time of about 100 ps, well separated from the water
relaxation time (approximately 10 ps).  On the other hand, these
results show that kinetic contributions can lead not just to the
Hubbard-Onsager decrement, but also to kinetic \emph{increments},
which arise from the completely different mechanism of pairing.

The reliability of the KBFF results offers the opportunity to test
quantitatively the Hubbard-Onsager continuum
theory~\cite{hubbard77a,hubbard78a} and the Hubbard-Colonomos-Wolynes
molecular theory\cite{hubbard79a} for the kinetic decrement. Obtaining
a precise estimate of the kinetic decrement from the analysis of
the spectrum $C_{wi}(\omega)$, however, can be computationally quite
demanding, especially if the static contribution is small. In the
present case, in fact, the fluctuations of the spectrum of the
cross-correlation function at low frequencies did not allow us to
perform a reliable extrapolation to $\omega=0$. The interpretation
in terms of linear response of the correlation $C_{wi}$ suggests,
however, that it is possible to compute the decrement using a non-equilibrium MD (NEMD)
approach. This consists in applying a constant (fictitious) electric
field acting only on ions, and in measuring the induced polarization
in water by the ion flow. This method realizes direcly the thought
experiment proposed by Hubbard and coworkers\cite{hubbard79a} to
explain the first dynamic mechanism for the decrement.
Calculating the dielectric decrement using this NEMD approach has
obvious advantages over the linear-response approach: relatively
high fields can be used to increase the signal-to-noise ratio, and
no extrapolation procedure is needed.  Indeed, a 500~ps long run
with an electric field of about 0.25 V/nm is enough to determine
the dynamic contribution $\Delta\epsilon_{wi}=-1.7$ with an accuracy
of 5\%. This term is only half of the contribution to the total
dielectric decrement, which therefore amounts to $-3.4$. Two separate
simulations with a different applied voltage were used to to control
the validity of linear response assumption.

The other kinetic contribution to the static permittivity is due
to the ion-ion current correlation, but unlike the water-ion cross term this can not be computed
 using a NEMD approach,
since the ionic cloud is infinitely polarizable. As it is seen from
the spectrum of $C_{iT}$ in Fig.~\ref{fig:kbff}, the scatter of the
data at low frequencies is rather large. This makes a precise
extrapolation to zero frequency very difficult. However, a rough
estimate taking into account the negative $\Delta\epsilon_{wi}$
contribution  suggests a value of $\Delta\epsilon_{ii}$ of about
2-5.  Such a contribution would in large part counterbalance the
total kinetic decrement $2\Delta\epsilon_{wi}$.  

Coming back to the
ion-water contribution,  we now compare the simulation results
with the available theoretical models.  The continuum
theory\cite{hubbard77a,hubbard78a} requires as input parameters the
main dielectric relaxation time and the conductivity of the solution.
Using the values computed by the MD simulation, the continuum model
predicts for the KBFF case a considerably larger decrement
$\Delta\epsilon_{HO,mol}\simeq-8$ (to be compared with
$2\Delta_{wi}=-3.4$). The situation does not improve if the molecular
theory developed by Hubbard, Colonomos and Wolynes\cite{hubbard79a}
is used. For a system with many ions in solution the molecular theory
estimate of the dielectric decrement is
\begin{equation}
\Delta\epsilon_{HO,mol}=
-8\pi \frac{\tau_{wT}}{3 e} \avg{\sum_{ij} R_i \sigma_i  \bsym{\mu}_j \cdot
\vec{r}_{ij} / r_{ij}^3 },\label{eq:molecular}
\end{equation} where $\bsym{\mu}_j$ is the dipole
moment of the $j-$th water molecule, located at the relative position
$\vec{r}_{ij}$ from the $i$-th ion of radius $R_i$ and specific
conductance $\sigma_i$ (i.e., the contribution to the total conductance
stemming from the single ion current). Applying Eq.~(\ref{eq:molecular}) to
the 
KBFF case leads to $\Delta\epsilon_{HO,cont}\simeq-19$,
which is in absolute value even larger than $\Delta\epsilon_{HO,cont}$.
This is because $\Delta\epsilon_{HO,mol}$ is very sensitive to the
local structure of water around each ion, due to the $1/r_{ij}$
term in the statistical average.  

In conclusion, we showed that the phenomenon of kinetic decrement exists 
and can account  for a considerable
fraction of the static dielectric permittivity in salt solutions.
However,  both the the continuum and molecular theories for the dielectric
decrement fail describing the decrement at the present concentrations. The
reason for this failure has be sought in the inadequacy of the infinite
dilution approximation. The high salt density screens the electrostatic
interaction, therefore reducing the efficiency of the Hubbard-Onsager
mechanism.  This has been confirmed by a supplementary simulation at
lower concentration (60 ion pairs and same number of water molecules),
for which the kinetic contribution, $\Delta\epsilon_{wi}=-2.3$, is higher
(in absolute value), despite the lower conductivity of the solution. In
the limit of a continuum charged background, in fact, no Hubbard-Onsager
mechanism can take place. Our findings show also that a general class
of kinetic effects can be identified, whenever an observable can be
expressed as the time integral of a cross correlation. Quantities such
as electrophoretic mobility and viscosity all fall into this category,
and might possibly show similar kinetic contributions. Investigations
in this direction are ongoing.

\begin{acknowledgments}MS acknowledges support from the European
Community's Seventh Framework Programme (FP7-PEOPLE-2012-IEF) funded
under grant Nr.~331932 SIDIS.  SSK acknowledges support from
RFBR grants mol-a 1202-31-374 and mol-a-ved 12-02-33106, from the
Ministry of Science and Education of RF 2.609.2011 and, from Austrian
Science Fund (FWF): START-Projekt Y 627-N27. AA and CH acknowledge
support by the cluster of excellence SimTech of the University of
Stuttgart. The authors thank Friedrich Kremer, Christian
Schr\"oder and Othmar Steinhauser for useful discussions.

\end{acknowledgments}


\begin{thebibliography}{21}
\expandafter\ifx\csname natexlab\endcsname\relax\def\natexlab#1{#1}\fi
\expandafter\ifx\csname bibnamefont\endcsname\relax
  \def\bibnamefont#1{#1}\fi
\expandafter\ifx\csname bibfnamefont\endcsname\relax
  \def\bibfnamefont#1{#1}\fi
\expandafter\ifx\csname citenamefont\endcsname\relax
  \def\citenamefont#1{#1}\fi
\expandafter\ifx\csname url\endcsname\relax
  \def\url#1{\texttt{#1}}\fi
\expandafter\ifx\csname urlprefix\endcsname\relax\def\urlprefix{URL }\fi
\providecommand{\bibinfo}[2]{#2}
\providecommand{\eprint}[2][]{\url{#2}}

\bibitem[{\citenamefont{Kremer and Sch\"{o}nhals}(2003)}]{kremer03a}
\bibinfo{editor}{\bibfnamefont{F.}~\bibnamefont{Kremer}} \bibnamefont{and}
  \bibinfo{editor}{\bibfnamefont{A.}~\bibnamefont{Sch\"{o}nhals}}, eds.,
  \emph{\bibinfo{title}{Broadband Dielectric Spectroscopy}}
  (\bibinfo{publisher}{Springer}, \bibinfo{address}{Berlin},
  \bibinfo{year}{2003}).

\bibitem[{\citenamefont{Hubbard and Onsager}(1977)}]{hubbard77a}
\bibinfo{author}{\bibfnamefont{J.~B.} \bibnamefont{Hubbard}} \bibnamefont{and}
  \bibinfo{author}{\bibfnamefont{L.~J.} \bibnamefont{Onsager}},
  \bibinfo{journal}{J. Chem. Phys.} \textbf{\bibinfo{volume}{67}},
  \bibinfo{pages}{4850} (\bibinfo{year}{1977}).

\bibitem[{\citenamefont{Wolynes}(1980)}]{wolynes80a}
\bibinfo{author}{\bibfnamefont{P.~G.} \bibnamefont{Wolynes}},
  \bibinfo{journal}{Ann. Rev. Phys. Chem.} \textbf{\bibinfo{volume}{31}},
  \bibinfo{pages}{345} (\bibinfo{year}{1980}).

\bibitem[{\citenamefont{Caillol et~al.}(1986)\citenamefont{Caillol, Levesque,
  and Weis}}]{caillol86a}
\bibinfo{author}{\bibfnamefont{J.~M.} \bibnamefont{Caillol}},
  \bibinfo{author}{\bibfnamefont{D.}~\bibnamefont{Levesque}}, \bibnamefont{and}
  \bibinfo{author}{\bibfnamefont{J.~J.} \bibnamefont{Weis}},
  \bibinfo{journal}{J. Chem. Phys.} \textbf{\bibinfo{volume}{85}},
  \bibinfo{pages}{6645} (\bibinfo{year}{1986}).

\bibitem[{\citenamefont{Schr{\"o}der et~al.}(2006)\citenamefont{Schr{\"o}der,
  Rudas, and Steinhauser}}]{schroeder06c}
\bibinfo{author}{\bibfnamefont{C.}~\bibnamefont{Schr{\"o}der}},
  \bibinfo{author}{\bibfnamefont{T.}~\bibnamefont{Rudas}}, \bibnamefont{and}
  \bibinfo{author}{\bibfnamefont{O.}~\bibnamefont{Steinhauser}},
  \bibinfo{journal}{J. Chem. Phys.} \textbf{\bibinfo{volume}{125}},
  \bibinfo{pages}{244506} (\bibinfo{year}{2006}).

\bibitem[{\citenamefont{Sega et~al.}(2013)\citenamefont{Sega, Kantorovich,
  Arnold, and Holm}}]{sega13a}
\bibinfo{author}{\bibfnamefont{M.}~\bibnamefont{Sega}},
  \bibinfo{author}{\bibfnamefont{S.~S.} \bibnamefont{Kantorovich}},
  \bibinfo{author}{\bibfnamefont{A.}~\bibnamefont{Arnold}}, \bibnamefont{and}
  \bibinfo{author}{\bibfnamefont{C.}~\bibnamefont{Holm}}, in
  \emph{\bibinfo{booktitle}{Recent Advances in Broadband Dielectric
  Spectroscopy}}, edited by \bibinfo{editor}{\bibfnamefont{Y.~P.}
  \bibnamefont{Kalmykov}} (\bibinfo{publisher}{Springer},
  \bibinfo{address}{Dordrecht}, \bibinfo{year}{2013}), NATO Science Peace S.,
  chap.~\bibinfo{chapter}{8}, pp. \bibinfo{pages}{103--122}.

\bibitem[{\citenamefont{Kubo}(1957)}]{kubo57a}
\bibinfo{author}{\bibfnamefont{R.}~\bibnamefont{Kubo}}, \bibinfo{journal}{J.
  Phys. Soc. Jpn.} \textbf{\bibinfo{volume}{12}}, \bibinfo{pages}{570}
  (\bibinfo{year}{1957}).

\bibitem[{\citenamefont{Berendsen et~al.}(1987)\citenamefont{Berendsen,
  Grigera, and Straatsma}}]{berendsen87a}
\bibinfo{author}{\bibfnamefont{H.~J.~C.} \bibnamefont{Berendsen}},
  \bibinfo{author}{\bibfnamefont{J.~R.} \bibnamefont{Grigera}},
  \bibnamefont{and} \bibinfo{author}{\bibfnamefont{T.~P.}
  \bibnamefont{Straatsma}}, \bibinfo{journal}{J. Phys. Chem.}
  \textbf{\bibinfo{volume}{91}}, \bibinfo{pages}{6269} (\bibinfo{year}{1987}),
  ISSN \bibinfo{issn}{0022-3654}.

\bibitem[{\citenamefont{Scott et~al.}(1999)\citenamefont{Scott,
  H{\"u}nenberger, Tironi, Mark, Billeter, Fennen, Torda, Huber, Kr{\"u}ger,
  and van Gunsteren}}]{scott99a}
\bibinfo{author}{\bibfnamefont{W.~R.~P.} \bibnamefont{Scott}},
  \bibinfo{author}{\bibfnamefont{P.~H.} \bibnamefont{H{\"u}nenberger}},
  \bibinfo{author}{\bibfnamefont{I.~G.} \bibnamefont{Tironi}},
  \bibinfo{author}{\bibfnamefont{A.~E.} \bibnamefont{Mark}},
  \bibinfo{author}{\bibfnamefont{S.~R.} \bibnamefont{Billeter}},
  \bibinfo{author}{\bibfnamefont{J.}~\bibnamefont{Fennen}},
  \bibinfo{author}{\bibfnamefont{A.~E.} \bibnamefont{Torda}},
  \bibinfo{author}{\bibfnamefont{T.}~\bibnamefont{Huber}},
  \bibinfo{author}{\bibfnamefont{P.}~\bibnamefont{Kr{\"u}ger}},
  \bibnamefont{and} \bibinfo{author}{\bibfnamefont{W.~F.} \bibnamefont{van
  Gunsteren}}, \bibinfo{journal}{J. Phys. Chem. A}
  \textbf{\bibinfo{volume}{103}}, \bibinfo{pages}{3596} (\bibinfo{year}{1999}).

\bibitem[{\citenamefont{Weerasinghe and Smith}(2003)}]{weerasinghe03a}
\bibinfo{author}{\bibfnamefont{S.}~\bibnamefont{Weerasinghe}} \bibnamefont{and}
  \bibinfo{author}{\bibfnamefont{P.~E.} \bibnamefont{Smith}},
  \bibinfo{journal}{J. Chem. Phys.} \textbf{\bibinfo{volume}{119}},
  \bibinfo{pages}{11342} (\bibinfo{year}{2003}).

\bibitem[{\citenamefont{Nos\'{e}}(1984)}]{nose84a}
\bibinfo{author}{\bibfnamefont{S.}~\bibnamefont{Nos\'{e}}},
  \bibinfo{journal}{Molecular Physics} \textbf{\bibinfo{volume}{52}},
  \bibinfo{pages}{255} (\bibinfo{year}{1984}).

\bibitem[{\citenamefont{Hoover}(1985)}]{hoover85a}
\bibinfo{author}{\bibfnamefont{W.~G.} \bibnamefont{Hoover}},
  \bibinfo{journal}{Phys. Rev. A} \textbf{\bibinfo{volume}{31}},
  \bibinfo{pages}{1695} (\bibinfo{year}{1985}).

\bibitem[{\citenamefont{Parrinello and Rahman}(1981)}]{parrinello81a}
\bibinfo{author}{\bibfnamefont{M.}~\bibnamefont{Parrinello}} \bibnamefont{and}
  \bibinfo{author}{\bibfnamefont{A.}~\bibnamefont{Rahman}},
  \bibinfo{journal}{J. Appl. Phys.} \textbf{\bibinfo{volume}{52}},
  \bibinfo{pages}{7182} (\bibinfo{year}{1981}).

\bibitem[{\citenamefont{Essmann et~al.}(1995)\citenamefont{Essmann, Perera,
  Berkowitz, Darden, Lee, and Pedersen}}]{essmann95a}
\bibinfo{author}{\bibfnamefont{U.}~\bibnamefont{Essmann}},
  \bibinfo{author}{\bibfnamefont{L.}~\bibnamefont{Perera}},
  \bibinfo{author}{\bibfnamefont{M.~L.} \bibnamefont{Berkowitz}},
  \bibinfo{author}{\bibfnamefont{T.}~\bibnamefont{Darden}},
  \bibinfo{author}{\bibfnamefont{H.}~\bibnamefont{Lee}}, \bibnamefont{and}
  \bibinfo{author}{\bibfnamefont{L.}~\bibnamefont{Pedersen}},
  \bibinfo{journal}{J. Chem. Phys.} \textbf{\bibinfo{volume}{103}},
  \bibinfo{pages}{8577} (\bibinfo{year}{1995}).

\bibitem[{\citenamefont{Futurelle and McGinty}(1971)}]{futurelle71a}
\bibinfo{author}{\bibfnamefont{R.~P.} \bibnamefont{Futurelle}}
  \bibnamefont{and} \bibinfo{author}{\bibfnamefont{D.~J.}
  \bibnamefont{McGinty}}, \bibinfo{journal}{Chem. Phys. Lett.}
  \textbf{\bibinfo{volume}{12}}, \bibinfo{pages}{285} (\bibinfo{year}{1971}).

\bibitem[{\citenamefont{Allen and Tildesley}(1987)}]{allen87a}
\bibinfo{author}{\bibfnamefont{M.~P.} \bibnamefont{Allen}} \bibnamefont{and}
  \bibinfo{author}{\bibfnamefont{D.~J.} \bibnamefont{Tildesley}},
  \emph{\bibinfo{title}{Computer Simulation of Liquids}}, Oxford Science
  Publications (\bibinfo{publisher}{Clarendon Press},
  \bibinfo{address}{Oxford}, \bibinfo{year}{1987}), \bibinfo{edition}{1st} ed.

\bibitem[{\citenamefont{Press et~al.}(1992)\citenamefont{Press, Teukolsky,
  Vetterling, and Flannery}}]{press92a}
\bibinfo{author}{\bibfnamefont{W.~H.} \bibnamefont{Press}},
  \bibinfo{author}{\bibfnamefont{S.~A.} \bibnamefont{Teukolsky}},
  \bibinfo{author}{\bibfnamefont{W.~T.} \bibnamefont{Vetterling}},
  \bibnamefont{and} \bibinfo{author}{\bibfnamefont{B.~P.}
  \bibnamefont{Flannery}}, \emph{\bibinfo{title}{Numerical Recipes: The Art of
  Scientific Computing}} (\bibinfo{publisher}{Cambridge University Press},
  \bibinfo{address}{Cambridge}, \bibinfo{year}{1992}), \bibinfo{edition}{2nd}
  ed.

\bibitem[{\citenamefont{Hubbard et~al.}(1979)\citenamefont{Hubbard, Colonomos,
  and Wolynes}}]{hubbard79a}
\bibinfo{author}{\bibfnamefont{J.~B.} \bibnamefont{Hubbard}},
  \bibinfo{author}{\bibfnamefont{P.}~\bibnamefont{Colonomos}},
  \bibnamefont{and} \bibinfo{author}{\bibfnamefont{P.~G.}
  \bibnamefont{Wolynes}}, \bibinfo{journal}{J. Chem. Phys.}
  \textbf{\bibinfo{volume}{71}}, \bibinfo{pages}{2652} (\bibinfo{year}{1979}).

\bibitem[{\citenamefont{Hess et~al.}(2006)\citenamefont{Hess, Holm, and van~der
  Vegt}}]{hess06b}
\bibinfo{author}{\bibfnamefont{B.}~\bibnamefont{Hess}},
  \bibinfo{author}{\bibfnamefont{C.}~\bibnamefont{Holm}}, \bibnamefont{and}
  \bibinfo{author}{\bibfnamefont{N.}~\bibnamefont{van~der Vegt}},
  \bibinfo{journal}{J. Chem. Phys.} \textbf{\bibinfo{volume}{124}},
  \bibinfo{pages}{164509} (\bibinfo{year}{2006}).

\bibitem[{\citenamefont{Buchner}(20048)}]{buchner08a}
\bibinfo{author}{\bibfnamefont{R.}~\bibnamefont{Buchner}},
  \bibinfo{journal}{Pure Appl. Chem.} \textbf{\bibinfo{volume}{80}},
  \bibinfo{pages}{1239} (\bibinfo{year}{20048}).

\bibitem[{\citenamefont{Hubbard}(1978)}]{hubbard78a}
\bibinfo{author}{\bibfnamefont{J.~B.} \bibnamefont{Hubbard}},
  \bibinfo{journal}{J. Chem. Phys.} \textbf{\bibinfo{volume}{68}},
  \bibinfo{pages}{1649} (\bibinfo{year}{1978}).

\end{thebibliography}
\end{document}